# Drumhead Surface States of Rhombohedral Graphite with Near Ideal Quantum Geometry Condition

Lin-Lin Wang[1,2*]

[1]Ames National Laboratory, Ames, IA 50011, USA

[2]Department of Physics and Astronomy, Iowa State University, Ames, IA 50011, USA

*llw@ameslab.gov

# Abstract

The ideal quantum geometry (IQG) condition for the trace of quantum metric ($TrG(\boldsymbol{k})$) being equal to the magnitude of Berry curvature ($|\Omega(\boldsymbol{k})|$) as in $|\Omega|/TrG$=1 has been associated with realizing fractional Chern insulators for the flat bands in a few layers of rhombohedral graphene (RG). More recently the IQG condition has also been proposed for superconductivity in the flat bands of thick RG layers. Using density functional theory and Wannier functions, we study the symmetry-protected topology of bulk RG, drumhead surface states (DSS) of semi-infinite RG surface and slabs, and the IQG condition for thin and thick RG slabs. We find that bulk RG is a weak topological insulator with spin-orbit coupling (SOC), besides being effectively a chiral semimetal with chiral nodal lines without SOC. We also find that the DSS flat bands of thick RG slabs have a sizable convex curvature with dispersion depth agreeing with the recent angle-resolved photoemission spectroscopy experiment. The calculated IQG also shows a convex shape with the $K$ point at the center having a minimum with approximately $|\Omega|/TrG \approx 1$. But the inner rim of the DSS region approaches the IQG condition of $|\Omega|/TrG = 1$. These results on thick RG slabs from first-principles calculations provide useful information for the pristine DSS at the single-particle level for future studies to consider when enhanced electron correlation and many-body interaction will be included.

## I. Introduction

The 2-dimesional (2D) systems of stacking and twisting a few exfoliated van der Waals (vdW) layers have become versatile platforms to fine tune electron interaction and correlation to realize novel quantum states[1-5]. Among them, few-layer rhombohedral graphene (RG)[6-13] in tri-, tetra- and penta-layers present a catalog of topological and correlated quantum states from superconductivity (SC)[14-16], Mott insulator[17], Chern insulator[18, 19], half-metal[20], orbital multiferroicity[21] to fractional Chern insulator (FCI)[22, 23], which realizes fractional quantum anomalous Hall effect (FQAHE). The high tunability of electron interaction and correlation in RG comes from the flat bands around the *K*-points which can be modulated with interfacial moiré and electric displacement field. It is widely accepted that RG is a chiral semimetal with the drumhead surface states (DSS) flat bands at the Fermi energy ($E_F$) from the chiral nodal lines around the *K*-points[6, 24]. However, just as being a Dirac semimetal, graphene[25] is also a quantum spin Hall insulator (QSH)[26, 27] or 2D topological insulator (TI) if the small band gap (~μeV) from spin-orbit coupling (SOC) is considered, because the band degeneracy at the *K*-points is not protected by symmetry with SOC. Similarly, bulk graphite itself is also a high-order TI[28]. In this regard, it can be asked what symmetry-protected topological (SPT) state exists in bulk RG at the single-particle level, in addition to the opposite chirality-based valley Chern number at *K* and −*K*, which cancels out when summed over the whole Brillouin zone (BZ).

More recently, besides Berry curvature ($\Omega(\boldsymbol{k})$) as the imaginary part of the quantum geometry tensor (QGT), quantum metric ($G(\boldsymbol{k})$) as the real part of QGT[29-32] has been found to also play prominent roles in topology related properties including nonlinear transport[33] and optical properties[34]. Specifically, for the flat Chern bands tuned in few-layer RG, the ideal quantum geometry (IQG) condition for the trace of quantum metric being equal to the magnitude of Berry curvature ($|\Omega|/TrG$=1) has been found to help stabilize the fractional filling[35-42] as a special case of vortexable[43] Chern bands to realize FCI[44-48] for FQAHE and more. On the other hand, the IQG condition has also been proposed for thick layers of RG slab[49] to realize SC with the DSS flat bands, where quantum metric gives effective superfluid density despite of zero kinetic energy[50, 51]. However, a recent angle-resolved photoemission spectroscopy (ARPES) experiment has shown that the DSS of semi-infinite RG is not strictly flat but with a curvature of 32 meV[52], in contrast to the flatter band in penta-layer RG[53]. Such curvature in the DSS of RG has been noted before in earlier DFT calculations[10, 54] and also was proposed to facilitate SC with annular Fermi surface[55-57] as trigonal warping. But how the curvature of pristine DSS affects the IQG condition for thick RG slabs has not been well studied, before the subsequent enhanced electron correlation and many-body interaction are introduced via doping and electric displacement field.

Here using density functional theory (DFT)[58, 59] and Wannier functions[60-62], we study the SPT of bulk RG, DSS of semi-infinite RG surface and IQG condition for thick RG slabs. We find that bulk RG is a weak TI with SOC, besides being effectively a chiral semimetal with chiral nodal lines without SOC. We also find the DSS flat bands of semi-infinite and thick RG slabs have a sizable convex curvature agreeing with the recent ARPES[52]. The calculated IQG for the DSS of thick RG slabs also shows a convex shape, with the $K$-point at the center having a minimum of $|\Omega|/TrG$~0.96 instead of 1. But the inner rim of the DSS region shows a more ideal condition of $|\Omega|/TrG$=1. These results on thick RG slabs from first-principles calculation provide some missing but useful information of the pristine DSS flat bands at the single-particle level for the future studies to consider when enhanced electron correlation and many-body interaction are included.

## II. Computational Methods

Band structures of RG in space group 166 (R−3m) without and with SOC have been calculated in DFT with a plane-wave basis set and projected augmented wave method[63] as implemented in VASP[64, 65]. Exchange-correlation (XC) functional of PBE[66] with vdW correction of D3 and r2SCAN+rVV10[67] meta-GGA XC functional have been used. We have employed a Monkhorst-Pack[68] (15×15×3) $k$-point mesh with a Gaussian smearing of 0.05 eV including the $\Gamma$ point and a kinetic energy cutoff of 400 eV. The experimental structural parameters[69] of $a$=2.461 Å and $c$=10.061 Å have been used for all the band structure calculations. To explore the surface states and band structure topological properties, Wannier functions[60-62] have been constructed to reproduce closely the bulk band structure including SOC in the range of $E_F$±3eV with C *sp* orbitals. The Wilsom loop, slab band structures and surface spectral functions with the surface Green's function methods[70, [71] have been calculated in WannierTools[72]. For the quantum geometry tensor calculation in 2D slabs, we have used the Kubo style formula

$$\Omega(\boldsymbol{k}) = -2\, Im \sum_{n}\sum_{m} \frac{v_{nm,x}(\boldsymbol{k}) v_{mn,y}(\boldsymbol{k})}{[\omega_m(\boldsymbol{k}) - \omega_n(\boldsymbol{k})]^2} \tag{1}$$

$$\mathrm{G}(\boldsymbol{k}) = Re \sum_{n}\sum_{m} \frac{\left[v_{nm,x}(\boldsymbol{k}) v_{mn,x}(\boldsymbol{k}) + v_{nm,y}(\boldsymbol{k}) v_{mn,y}(\boldsymbol{k})\right]}{[\omega_m(\boldsymbol{k}) - \omega_n(\boldsymbol{k})]^2} \tag{2}$$

where $n$ and $m$ run through valence and conduction bands, respectively. The $\omega_n(\boldsymbol{k}) = \varepsilon_n(\boldsymbol{k})/\hbar$ are the energy eigenvalues, and the velocity matrix elements $v_{nm,\alpha}(\boldsymbol{k})$ are calculated from the Hamiltonian $\widehat{H}(k)$ as

$$v_{nm,\alpha}(\boldsymbol{k}) = \frac{1}{\hbar}\left\langle u_{n\boldsymbol{k}} \middle| \frac{\partial \widehat{H}(\boldsymbol{k})}{\partial k_\alpha} \middle| u_{m\boldsymbol{k}} \right\rangle \tag{3}$$

where $\alpha = x$ or $y$, and $u_{n\boldsymbol{k}}$ are the Kohn-Sham single-particle wavefunctions.

## III. Results and Discussion

Figure 1 presents the bulk crystal and band structures of RG. The rhombohedral crystal structure in space group 166 (R–3m) with ABC stacking of honeycomb layers is shown in Fig.1(a), which has a 3-fold rotation axis, inversion symmetry, a set of vertical mirror planes combined with 2-fold rotation axis. The primitive unit cell of RG in Fig.1(a) has a single honeycomb layer with two C atoms, while the conventional hexagonal unit cell has three honeycomb layers with six C atoms. The vdW interlayer distance of 3.354 Å in RG[69] is 3.5% smaller than the 3.473 Å in graphite[73]. The bulk band structure in Fig.1(c) shows RG is a semimetal without SOC, having a large band gap along most of high-symmetry directions, except for around the $K$-point of the Brillouin zone (BZ), which is projected to the $\bar{K}$ point of the hexagonal 2D BZ in Fig.1(b). The band crossings around the $K$-point form a spiral nodal line along the $k_z$ direction as plotted in Fig.1(d) with one of points $K_1$ being used in the band structure plot in Fig.1(c). When projected on the (001) plane in Fig.1(e), the chiral nodal line forms a 2D nodal loop around the $\bar{K}$ point, which has a warped trigonal shape from the 3-fold rotation. This spiral nodal line is gapped out when SOC is turned on, which results in a very small band gap of 0.03 meV due to the small SOC strength of C atoms.

The band structures of stacked few layers RG without SOC are plotted in Fig.1(f-i) for L=3, 4, 5 and 6, respectively, which agree well with previous tight-binding (TB)[7-9] and DFT studies[10, 54, 74]. For L=3, there is a chiral Dirac crossing point just off the *K* point. For L=4, 5 and 6, the degenerated flat valence and conduction bands grow in range toward the chiral nodal line boundary (outer vertical dashed lines) from the band inversion. As the range of flat bands grows with L, these bands around the *K* point also become less flat, which is more clearly seen in Fig.1(j) for L=18 because of the more interlayer interactions along the *z* direction. For such a thick RG slab, the bulk bands form a band inversion region as bounded by the chiral nodal line projection. The DSS flat bands reside inside this band inversion region. Noticeably the DSS is not strictly flat, but has a convex shape with a depth of 20 meV, which is in line with the recent ARPES measurement of 32 meV on semi-infinite RG (001) surface[52].

To better depict the topological surface states of RG, we have calculated the surface spectral functions for semi-infinite RG in Fig.2 with an increased SOC factor of 100, which does not change the band order or topological index. First as plotted in Fig.2(a-b) for the Wilson loop of the Wannier charge center (WCC) on the $k_z$=0.0 and 0.5 planes, both are topologically non-trivial with an odd winding number. This corresponds to $Z_{2w3}$=1 and $Z_4$=0 in RG for a weak TI, which can be understood as the stacking of the QSH of graphene layers. Next on the side surface with broken C-C bonds, the surface states in Fig.2(c) shows two surface Dirac points around $E_F-1.5$ eV at time-reversal invariant momentum (TRIM) of $\bar{\Gamma}$ and $\bar{Y}$ points as protected by time-reversal symmetry (TRS). The even number of surface Dirac points also confirms the weak TI bulk-boundary correspondence. In contrast, on the naturally cleaving (001) surface as shown in Fig.2(d), there is no surface Dirac point at the TRIM of $\bar{\Gamma}$ and $\bar{M}$ points. The high intensity around the $\bar{K}$ point at $E_F$ is from the DSS as zoomed in Fig.2(e), which is comparable to the thick slab band structure in Fig.1(j). But here with SOC, the DSS is gapped, which can be more clearly seen in Fig.2(f) with surface-only contributions, because RG is a weak TI and $\bar{K}$ is not a TRIM point.

It is noticeable that the DSS for semi-infinite RG (001) in Fig.2(f) is not flat and more curved with a depth of 38 meV agreeing better with the ARPES measurement of 32 meV[52] than the 20 meV of L=18 in Fig.1(j). This is because as the layer number increases, more interlayer interactions are included along the *z* direction. As explained in Ref.[49], the strict flatness of DSS is met only when only the nearest-neighbor (NN) and next nearest-neighbor (NNN) hoping terms are used in the TB model, while here using Wannier functions, more hopping terms with longer range are automatically included. These almost flat DSS of RG have been proposed[49] to host the IQG condition, $|\Omega|/TrG$=1, with equal magnitude of Berry curvature and trace of quantum metric around the $\bar{K}$ point for DSS. Next, we will calculate the quantum geometry tensor of the DSS according to Eqn.(1-3) to see how IQG behave for the pristine DSS without doping.

For calculating the quantum geometry tensor of the DSS, we have used the RG slabs with increasing number of layers built with bulk RG Wannier Hamiltonian. The slab band structure of L=60 with normal SOC strength is plotted in Fig.3(a), the DSS with high projection on surface is clearly shown inside the band inversion region centered at the $K$ point agreeing with those in Fig.2(e) of the semi-infinite surface, albeit with a much-reduced gap between DSS valence and conduction bands. The DSS curved depth of 34 meV is also smaller than the 38 meV for the semi-infinite surface. The ratio of $|\Omega|/TrG$ for L=9, 18 and 60 are plotted in Fig.3(b). The trend of the $|\Omega|/TrG$ profiles agree with those in Ref.[49] (the number of hexagonal unit cells 3N=L). For the thin slab of L=9, the trace condition of $|\Omega|/TrG$ has a large variation. But for L=18, the trace condition already has the convex profile with slightly smaller value of $|\Omega|/TrG$ at the center region of $K$-point than the rim region. For the thick slab of L=60, the rim expands as indicated by the boundaries of $|\Omega|/TrG$ and its value approaches the IQG condition of 1. In contrast, at the center of the $K$-point, $|\Omega|/TrG$=0.960 with a converged convex profile.

To assess the effect of different XC functional on the DSS and quantum geometry, we have used r2SCAN+rVV10 and plotted the results in Fig.3(c-d). The inverted band gap at the $\bar{K}$ point, or separation between the bulk valence and conduction bands increases because meta-GGA tends to correct the underestimation of band gap from electron self-interaction in PBE. But the size of the band inversion region with DSS stays almost the same because the same lattice constants are used. As the results, the curvature of the DSS in Fig.3(c) has a slightly larger depth of 36 meV than the 34 meV in Fig.3(a). For the $|\Omega|/TrG$ in Fig.3(d), except for L=9 having a different variation, the converged profile for thicker L=18 and 60 stays the same with the central $|\Omega|/TrG$ being slightly reduced to 0.958 when compared to the 0.960 in Fig.3(b) calculated with PBE. This shows the robustness of our calculated results for $|\Omega|/TrG$ and the IQG of $|\Omega|/TrG$=1 is more fulfilled at the rim region than the center at the *K*-point.

For a closer look at the quantum geometry tensor, we plot separately $|\Omega|$ and $TrG$ in Fig.4(a). As expected, both have prominent peaks at the DSS boundaries where the flat surface states merge into much more dispersive bulk bands. The $TrG$ peaks have larger width and value than the corresponding peaks of $|\Omega|$. The larger width is from the outer rim, while the two are closely aligned at the inner rim and center, giving the near IQG condition inside the DSS region. To study the different behavior of the different components of quantum geometry tensor, we have plotted the $xx$ and $yy$ component of the quantum metric by zooming in the left and right DSS rims in Fig.4(b) and (c), respectively. Comparing to the off-diagonal $|\Omega|$, the trace of quantum metric $TrG$ has an earlier and more gradual buildup when going across the DSS rim. Among the $TrG$ components, $G_{xx}$ is larger than $G_{yy}$, while $G_{yy}$ has a shaper resonant with $|\Omega|$ than $G_{xx}$. This difference stems from the inequivalent in-plane directions with the *x*-axis pointing along the C nearest neighbor, while the *y*-axis along the C next nearest neighbor direction. Overall, inside the DSS rim, the inner rim has a sharper edge where $|\Omega|$ and $TrG$ are more closely aligned together to approach the IQG condition when compared to the outer rim. Such behavior can also be seen in the 2D plots of $|\Omega|$ and TrG in Fig.5. As zoomed in Fig.4(d) close to the center of the DSS interior region, both $TrG$ and $|\Omega|$ decrease moving toward the *K*-point with the minimum values at the center. But $|\Omega|$ decreases faster than $TrG$, giving a smaller ratio of $|\Omega|/TrG$ toward the center and thus the convex shape with near IQG. The two of them are closer at +0.03 1/Å toward the *M* direction than −0.03 1/Å toward the *Γ* direction because of the triangular not circular shape of the nodal line projection.

To zoom out and show the rim region with IQG condition more clearly in 2D, we have plotted the 2D quantum geometry tensor in Fig.5. Comparing to $|\Omega|$ in Fig.5(a), $TrG$ in Fig.5(b) has a more extended area on the outside of the band inversion region, agreeing with the line cut in Fig.4(a). Near the other valley around $\overline{K'}$ point, they are related by inversion or time-reversal with respect to the $\overline{K}$ point, quantum geometry tensors behave similarly, both with a 30-degree rotation. These 2D plots in Fig.5 and detailed analysis in Fig.4 show that the IQG condition of $|\Omega|/TrG$=1 can be met more strictly at the inner rim region of DSS, while it is only approximately met at the center with a convex profile.

To study the effects of the range of Wannier Hamiltonian on the DSS dispersion and IQG condition, we have truncated the Wannier Hamiltonian from the range of 15th to a shorter range of 2nd nearest neighbor in terms of unit cells, where the unit cell at origin is the conventional hexagonal unit cell with three atomic layers of six C atoms and 24 orbitals. Thus, even with the range of 2nd nearest neighbor for the hexagonal unit cell, the truncated Wannier Hamiltonian still has a longer range than those typically used for phenomenological tight-binding models, such as the $\gamma_4$, giving the finite dispersion of DSS (see Fig.S4 in Ref.[49]). As overlayed in Fig.6(a) and zoomed in Fig.6(b), the truncated Wannier Hamiltonian gives a smaller DSS region, and the dispersion is slightly flatter than the full Wannier Hamiltonian. But the IQG condition becomes more ideal in the smaller DSS region as plotted in Fig.6(c). The $|\Omega|/TrG$ at the *K*-point center is above 0.990 as zoomed in Fig.6.(d), but still has the convex shape. This behavior is also seen in the *TrG* plot in Fig.6(e), where $TrG$ peaks of the truncated Hamiltonian are sharper and narrower in spanning a smaller DSS region than the full Hamiltonian. However, the more ideal and flatter for the ratio of $|\Omega|/TrG$ does not mean flatter and smaller quantum geometry tensor themselves near the *K*-point center. Actually, as zoomed in Fig.6(f), the $|\Omega|$ and *TrG* of the truncated Hamiltonian are larger and more curved around the *K*-point center in the DSS interior region. But the $|\Omega|$ and *TrG* of the truncated Hamiltonian are more closely aligned together to make the ratio of $|\Omega|/TrG$ closer to unity and the IQG condition than those of the full Hamiltonian. These different behaviors can be understood by inspecting Eqn.(1-2). The DSS band energy curvature is only one factor that enters Eqn.(1-2) as the denominator, which are the same for both $|\Omega|$ and *TrG*. More important are the nominators resulted from the changes of velocity matrix elements and their relative magnitudes, reflecting the wave function spread and character change when going from the bulk-like region in the interior of slab to the surface layer, which determine the different magnitudes of $|\Omega|$ and *TrG*.

Like a few layers of RG, electron doping for thick layers of RG can modify the properties of DSS. For example, DSS splitting has been observed in ARPES[52] with electron doping, where the lower DSS band gains an increasingly large parabolic dispersion with more electron doping, while the upper near flat DSS band is pinned at $E_F$. A recent STM study[75] has shown that thick layers of RG can also be hole-doped with extra features and splitting from stacking fault at domain wall boundary. Another STM study[76] has shown that when $E_F$ can be precisely dialed inside the DSS gap of thick layers of RG, enhanced electron correlation and many-body interaction can give rise to a quantum magnet. Here we have focused on the pristine DSS of RG without doping at the single-particle level. The calculated 38 meV for DSS depth of finite dispersion is in a good agreement to the 32 meV observed in ARPES[52] without doping. The real samples in experiment can have defects from different stacking faults near domain walls, enhanced electron correlation and many-body interaction that are not accounted for in the current single-particle DFT calculations with a pristine RG structure. We understand the challenge and limitation of the DFT single-particle approach for lacking the enhanced electron correlation when putting extra electron into the near flat DSS as in the electron doping effect observed in ARPES. As discussed in the Supporting Information with Fig.S1, simply adding extra electron in the single-particle DFT calculation results in a large parabolic dispersion being shifted to lower energy for the DSS bands, but without the correlation-induced DSS splitting as observed in ARPES[52]. Thus, more detailed studies of these effects are needed for future work.

## IV. Conclusion

In conclusion, using calculations based on density functional theory (DFT) and Wannier functions, we have studied the symmetry-protected topology of bulk rhombohedral graphite (RG), drumhead surface states (DSS) of semi-infinite RG surface, and also the ideal quantum geometry (IQG) condition for thick RG slabs with the trace of quantum metric ($G(\boldsymbol{k})$)) and magnitude of Berry curvature ($\Omega(\boldsymbol{k})$) being equal as in $|\Omega|/TrG$=1. We find that bulk RG is a weak TI with spin-orbit coupling (SOC), besides being effectively a chiral semimetal with chiral nodal lines without SOC. We also find that the DSS flat bands of semi-infinite and thick RG slabs have a sizable convex curvature with a dispersion depth agreeing well with the recent ARPES experiment[52]. The calculated IQG for thick RG slabs shows a convex shape with the $K$ point at the center having a minimum of 0.96 instead of 1. But the rim of the DSS region shows a stricter IQG condition of $|\Omega|/TrG$=1. These results on semi-infinite surface and thick RG slabs from first-principles calculation provide some missing but useful information of the DSS and IQG at the single-particle level other than the usual tight-binding models for the future studies to consider when enhanced electron correlation and many-body interaction will be included.

The IQG condition has been proposed to stabilize the fractional Chern insulator in a few layers of RG systems with electric displacement field serving as an efficient knob to both enhance the IQG condition and dial the factional filling of the flat band. Here for pristine thick slabs of RG at the single-particle level, besides the good agreement of the DSS with ARPES experiment[52], we show that fulfilling of IQG condition has a crossover from thin (~9) to thick (~18) RG layers without electron doping. It will be interesting to see if such a crossover can be measured and confirmed in experiment with layer-dependent study to gradually fulfill IQG condition by realizing flat band superconductivity and other strongly correlated phenomena in thick RG layers.

**Data Availability**: The data that support the findings of this study are openly available[77].

## Acknowledgements

The work on quantum geometry calculation was supported by the Center for the Advancement of Topological Semimetals, an Energy Frontier Research Center funded by the U.S. Department of Energy Office of Science, Office of Basic Energy Sciences through the Ames National Laboratory. The work on band structure calculation was supported by the U.S. Department of Energy Office of Science, Office of Basic Energy Sciences through the Ames National Laboratory, Division of Materials Sciences and Engineering. The Ames National Laboratory is operated for the U.S. Department of Energy by Iowa State University under Contract No. DE-AC02-07CH11358. Some of this research used resources of the National Energy Research Scientific Computing Center (NERSC), a DOE Office of Science User Facility.

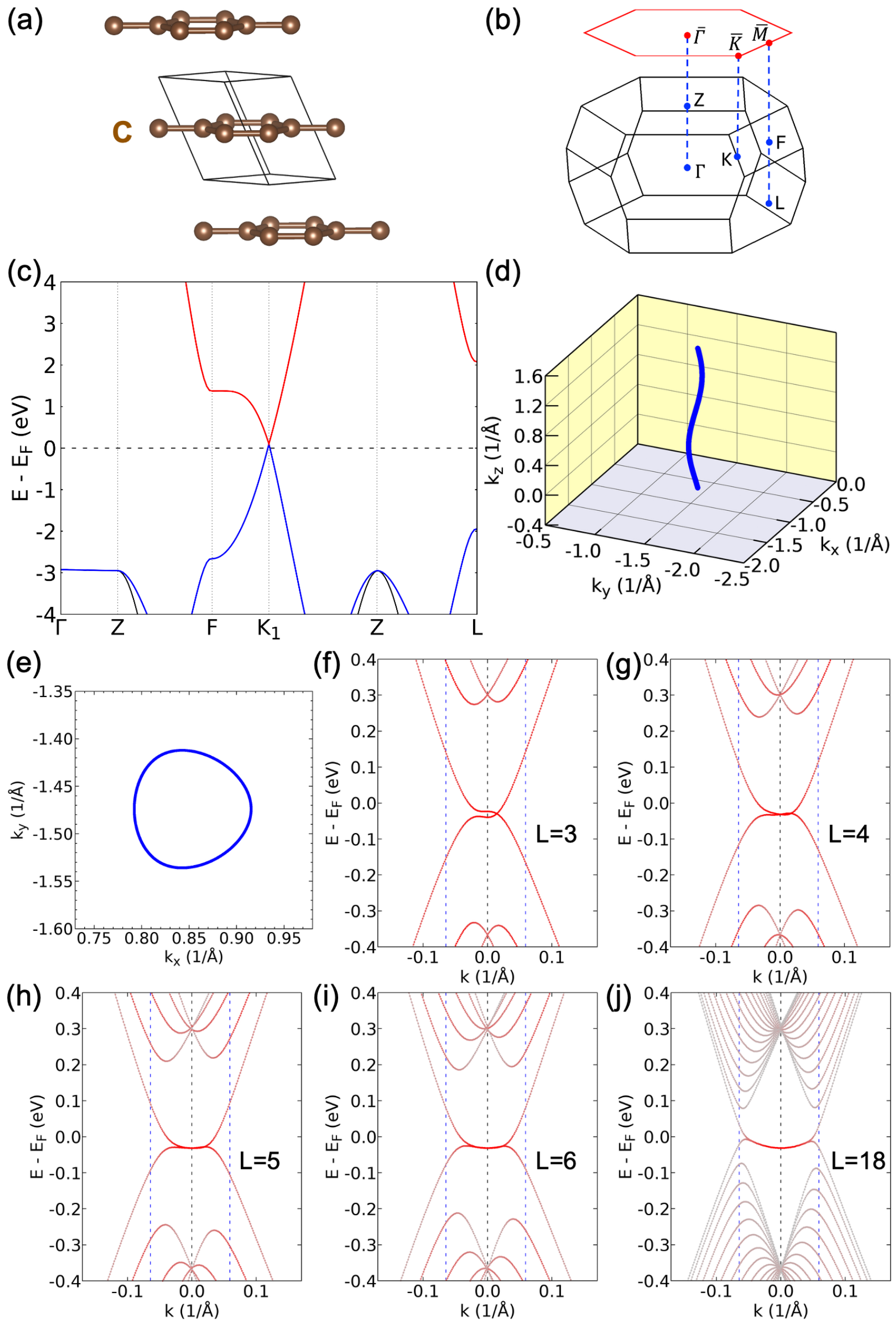

Figure 1. Band structure of rhombohedral graphite (RG) without spin-orbit coupling (SOC). (a) The crystal structure and primitive unit cell of RG with ABC stacking of honeycomb layers. (b) The 3D Brillouin zone (BZ) and (001) 2D surface BZ of RG with the high-symmetry $k$-points related via projection. (c) Band structure of RG along high-symmetry directions. The highest valence (lowest conduction) band is in blue (red). The $K_1$ point is one of the band crossing points near the $K$ point. (d) The chiral nodal line along the $k_z$ direction around the $K$ point. (e) The nodal loop around $\bar{K}$ on (001) as projected from the chiral nodal line. (f), (g), (h), (i) and (j) RG slab band dispersion around the $\bar{K}$ point along the $\bar{\Gamma}$-$\bar{K}$ direction for layer thickness (L) of 3, 4, 5, 6 and 18 as labeled. The outer blue vertical dashed lines show the projected nodal loop boundary.

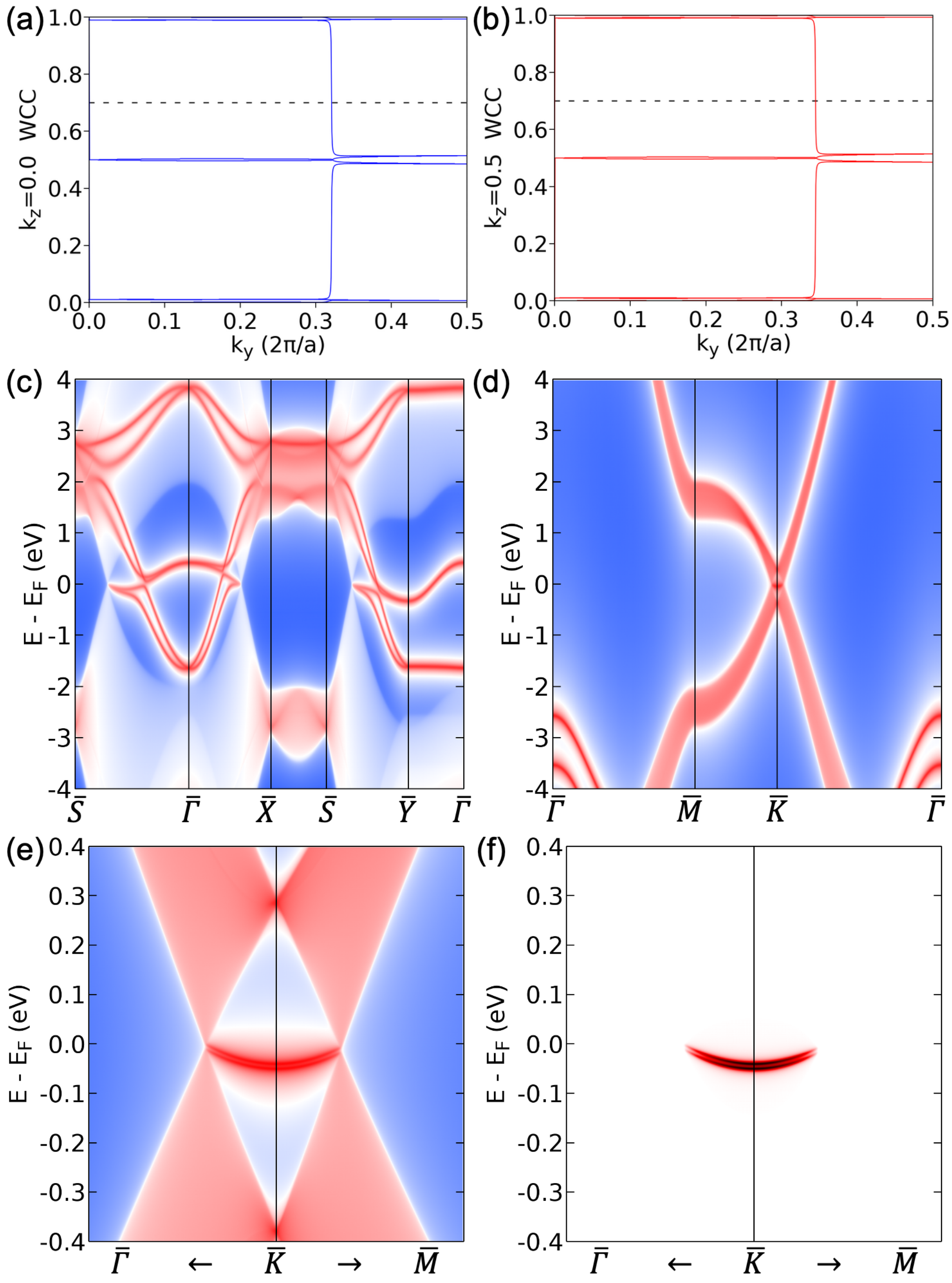

Figure 2. Topological surface band structure of rhombohedral graphite (RG) with enhanced spin-orbit coupling (SOC). Wilson loop of the Wannier charge center (WCC) on the (a) $k_z$=0.0 and (b) 0.5 planes. Surface spectral function on the (c) side and (d) (001) surface. (e) Surface spectral function zoomed around the (001) $\overline{K}$ point. (f) Same as (e) with surface-only contribution. The calculations are with a SOC factor of 100 to increase the surface band gap but with no change of band order or connectivity.

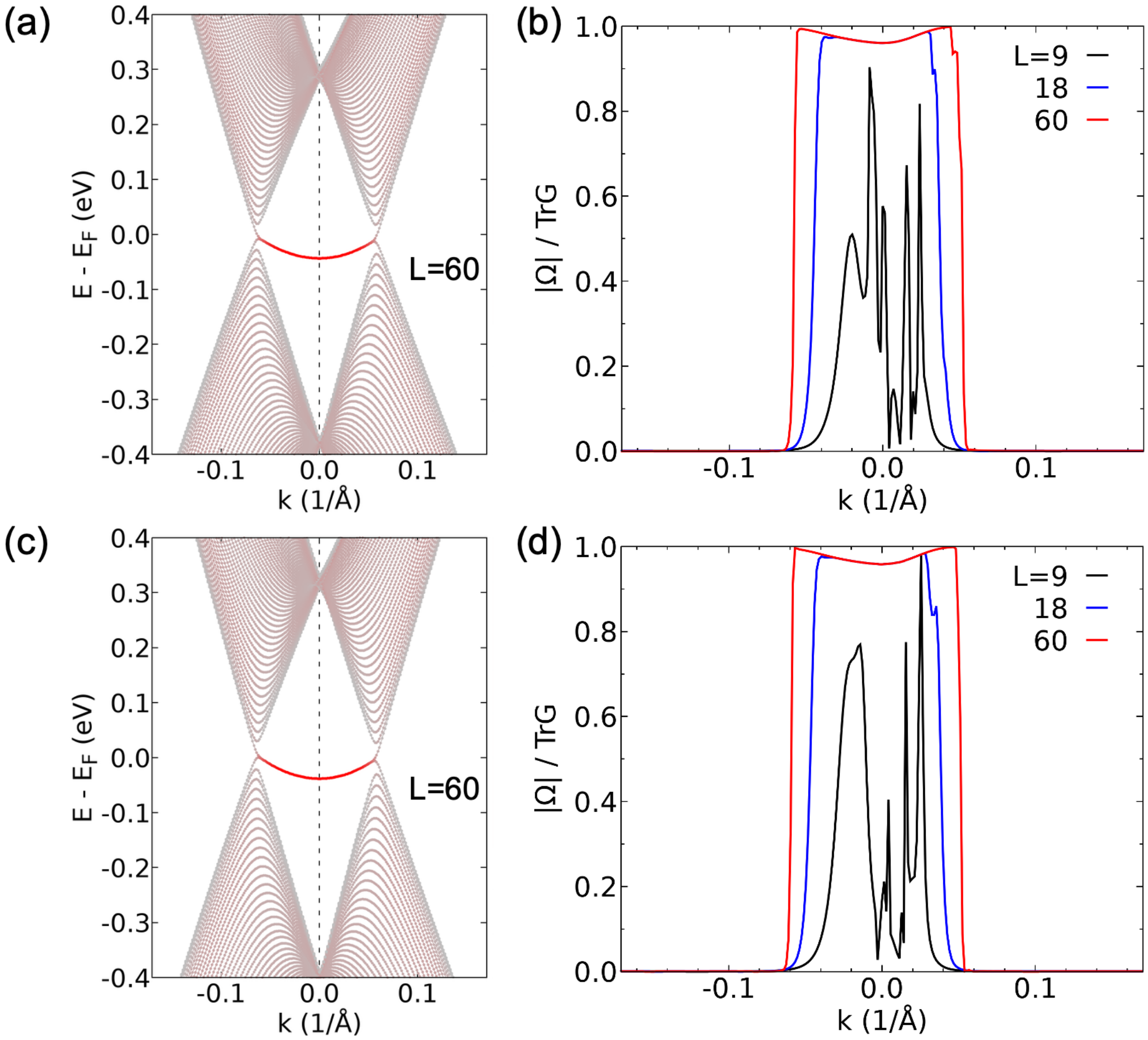

Figure 3. Quantum geometry of the drumhead surface states of rhombohedral graphite (RG). (a) Slab band structure of L=60 with the surface contribution shown in red. (b) Ratio between the magnitude of Berry curvature ($|\Omega|$) and trace of quantum metric ($TrG$) ($|\Omega|/TrG$) around the $\bar{K}$ point along the [110] direction for L=9, 18 and 60. (c-d) Same as (a-b) with r2SCAN+rVV10.

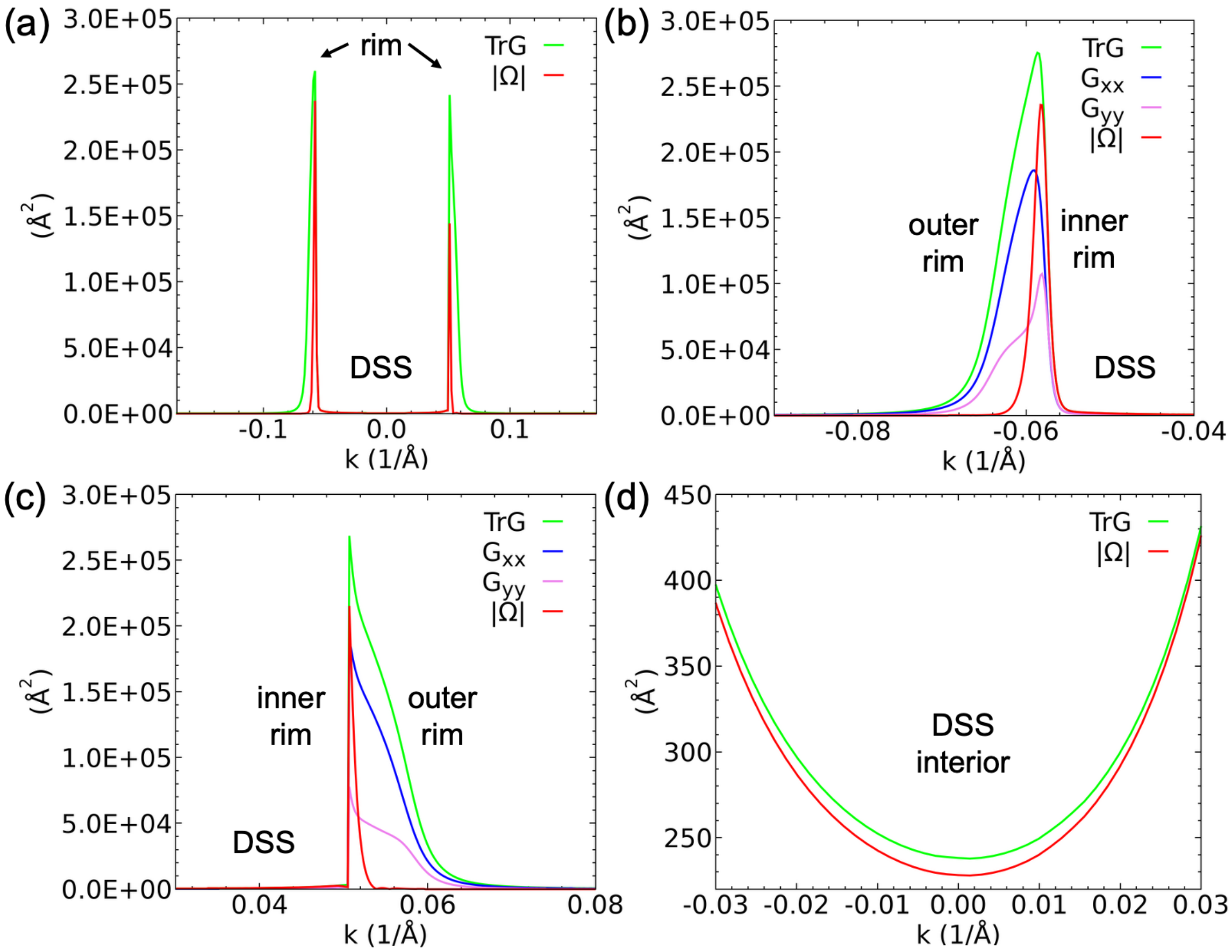

Figure 4. Quantum geometric tensor of the drumhead surface states (DSS) of rhombohedral graphite (RG). (a) Magnitude of Berry curvature (|Ω|) and trace of quantum metric (*TrG*) around the $\overline{K}$ point along the [110] direction for L=60. The boundary between the DSS region and the highly dispersive bulk band region is marked as rim, where quantum geometric tensor, |Ω| and *TrG*, have the sharp peaks. (b) Zoom-in of the left rim showing the relatively gradual rise of *TrG* and its components of $G_{xx}$ and $G_{yy}$ from the outer rim to inner rim edge in contrast to the much faster rise of |Ω|. (c) Similar to (b) zoom-in of the right rim. (d) Zoom-in of the DSS interior where *TrG* and |Ω| are closely aligned.

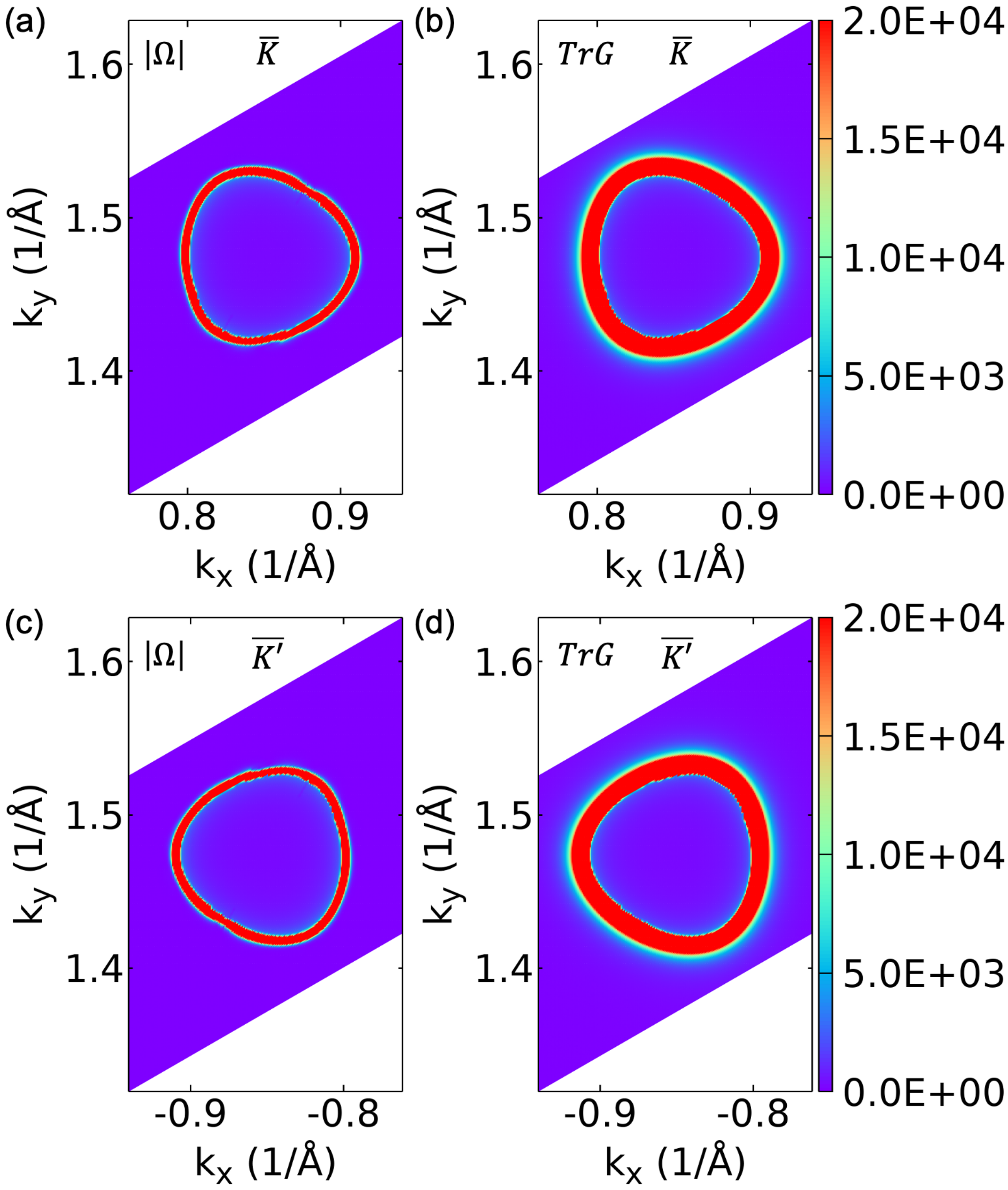

Figure 5. Quantum geometry of the drumhead surface states plotted in 2D for L=60. (a) Magnitude of Berry curvature (|Ω|) and (b) trace of quantum metric (TrG) around the $\overline{K}$ point calculated with r2SCAN+rVV10. (c-d) Same as (a-b) at the $\overline{K'}$ point.

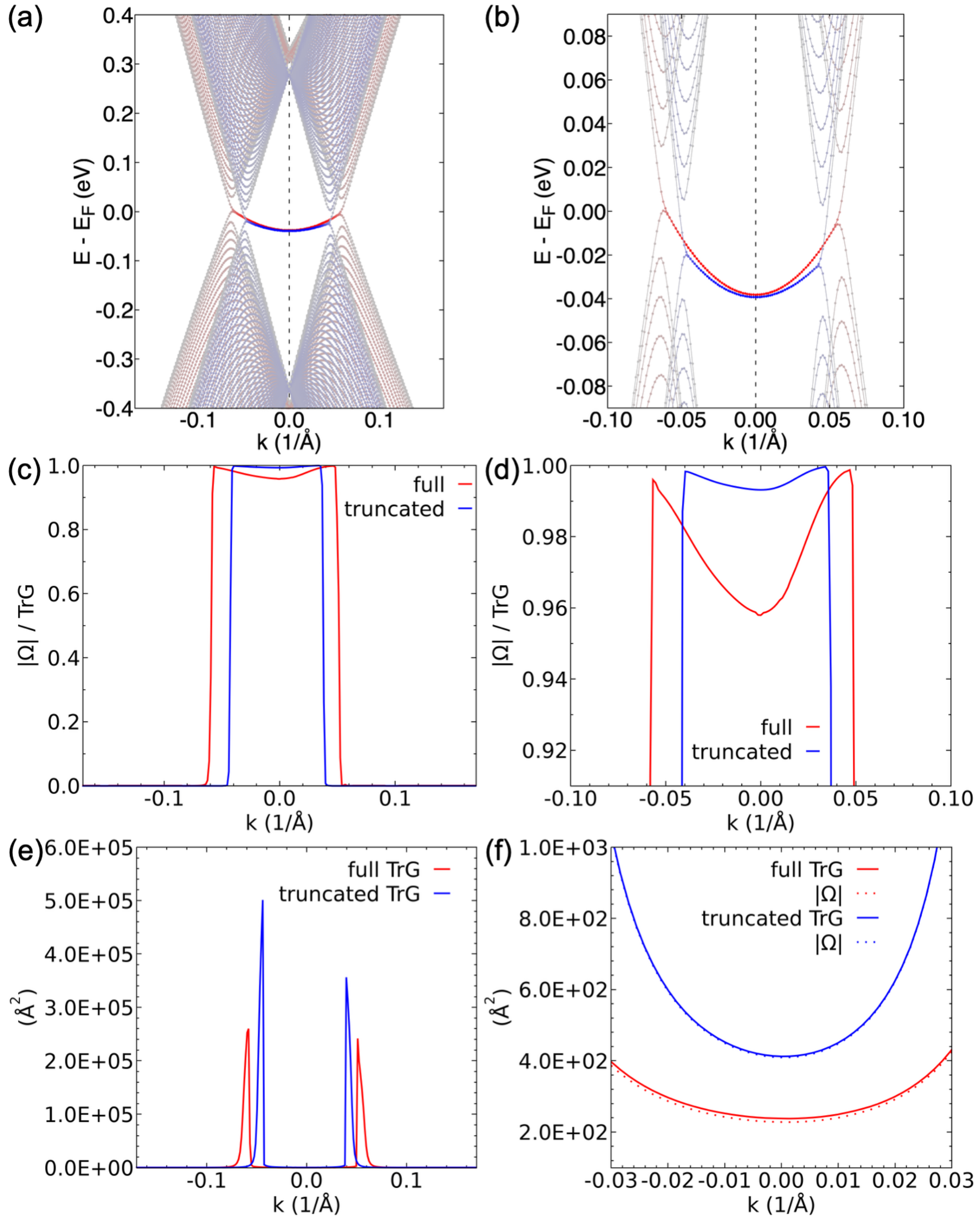

Figure 6. Comparison of quantum geometry for the drumhead surface states (DSS) of rhombohedral graphite (RG) between full and truncated Wannier Hamiltonians. (a) Overlayed slab band structures of L=60 with the DSS shown in red (blue) with the full (truncated) Hamiltonian around the $\overline{K}$ point along the [110] direction for L=60. (b) Zoom-in of (a) around the Fermi energy ($E_F$) showing the different scope and dispersion of the DSS with the full (red) vs. truncated (blue) Hamiltonian. (c) Ratio between the magnitude of Berry curvature ($|\Omega|$) and trace of quantum metric (*TrG*) and (d) its zoom-in. (e) *TrG* of the full vs. truncated Hamiltonian. (f) *TrG* and $|\Omega|$ zoomed in the DSS interior around the $\overline{K}$ point at $k$=0.00 for the full vs. truncated Hamiltonian.